\documentstyle[aps,multicol,epsfig]{revtex}
\newcommand{\nc}{\newcommand}
\nc{\be}{\begin{equation}}
\nc{\ee}{\end{equation}}
\nc{\bea}{\begin{eqnarray}}
\nc{\eea}{\end{eqnarray}}
\nc{\bean}{\begin{eqnarray*}}
\nc{\eean}{\end{eqnarray*}}
\nc{\mb}{\mbox}
\nc{\rnc}{\renewcommand}
\nc{\r}{\mb{\boldmath$r$}}
\nc{\x}{\mb{\boldmath$x$}}
\nc{\A}{\mb{\boldmath$A$}}
\nc{\sa}{\mb{\boldmath$a$}}
\nc{\nab}{\nabla}
\nc{\X}{\sf x}

\begin{document}
\draft

\def\del{\partial}

\title{ The Edge State Network Model and the Global Phase Diagram }
\author
{Kentaro Nomura and Daijiro Yoshioka
 }

\address{
Department of Basic Science, University of Tokyo, 3-8-1 Komaba, Tokyo 153-8902\\
}

\date{\today}
\maketitle

\begin{abstract} The effects of  randomness are investigated in the fractional quantum Hall systems.
Based on the Chern-Simons Ginzburg-Landau theory and considering relevant 
quasi-particle tunneling,
 the edge state network model for the hierarchical state is introduced 
and the plateau-plateau transition and the liquid-insulator 
transition are discussed.
This model has duality which corresponds to the relation of the 
quantum Hall liquid phase
 and 
the Hall insulating phase  and   reveals a mechanism of plateau
transition in the weak coupling regime.
\end{abstract}



\begin{multicols}{2}
\narrowtext

{\bf {1. Introduction}}

\vspace{3mm}
The integer and fractional quantum Hall effect appears 
in two-dimensional
 electron systems  in a strong magnetic field. 
For the quantum Hall effect weak randomness is needed. \cite{smg}
A large number of experimental and theoretical  
studies have been made 
on the effects of randomness. 
%
%
%
%
%
%
Especially, the quantum phase transition from a
quantum Hall liquid to a different quantum Hall liquid or 
the Hall insulating phase
which occurs as the randomness strength or the magnetic field
 increase are meaningful.\cite{gpda,gpdb,gpdc}
 In one experiment, Shahar et. al. 
observed a reflection symmetry in the $I-V_{xx}$ 
characteristics near the transition point of
 a $\nu=1/3$ fractional
 quantum Hall liquid to a Hall
insulator. \cite{shahar2} 
  They insisted  
that the observed symmetry corresponds to the charge-flux duality.
 \cite{scz}
Near the critical (i.e. self-dual) point the phase separation 
should occur,
Simshoni et al. and Pryadko et al.
introduced
the edge state network model and discussed the  non-linear 
$I-V_{xx}$ characteristics. \cite{simshoni,pryadko} 
The edge state network model was originally introduced by Chalker and Coddington,
to study the universal properties of the localization length in the strong magnetic
field in the framework of the non-interacting electron systems.\cite{cc}
Chalker-Coddington model is constructed by quantum tunneling of the electron near
the Fermi levels which are located along the equal potential lines.

 Meanwhile in the study of the liquid-liquid transition, 
the topological arguments and the numerical studies
have been successful in integer quantum Hall systems.\cite{gpd1,gpd2,gpd3,gpd4,gpd5,gpd6}
Using 
the flux attached composite fermion mean field theory  Kivelson, Lee, and Zhang(KLZ) mapped the fractional 
quantum Hall state to integer quantum Hall state and they 
proposed the global phase diagram.\cite{klz}
There are also experimental discussions on the validity of the global phase 
diagrams.\cite{ex1,ex1.1,ex2,ex2.2,ex3,ex4} 
Their results seem to be slightly different from KLZ's diagram.

In this paper 
we introduce a general fractional filling version of 
the network model. Our model shows dual symmetry which parallels the charge-flux
duality. In weak coupling regime this model describes the hierarchical quantum 
Hall liquid phase, whereas the strong coupling regime corresponds to the Hall 
insulator. Analyzing the model  the mechanisms of plateau transitions are revealed. We discuss
the global phase diagram finally.

\vspace{5mm}

{\bf {2. The edge state network model}}
\vspace{3mm}

The Chern-Simons gauge field theory is a long wave length and low energy effective 
theory of the fractional quantum Hall state.\cite{c-s1,scz}
In this theory one regards the electrons as the flux attached 
bosons.
 In the case of filling factor $\nu= 1/q$ ($q$  is an odd integer)  the Lagrangian is written as
\begin{eqnarray}
  {\cal L}[a_{\mu},\varphi ] &=&  -\frac{\nu}{4\pi}e^2 \ \epsilon^{\mu \nu \rho}a_{\mu}\partial_{\nu}a_{\rho}
  \nonumber \\ &&
  \ \ \ \ +   \overline{\varphi}[i\partial_0+e(a_0 + A_0)] \varphi \nonumber \\   &&
  \ \ \ \ \ \ \ \ \ +  \frac{1}{2m}|[-i\nab +e(\sa +\A)]\varphi|^2  \nonumber \\  &&
\ \ \ \ \ \ \ \ \ \ \   \ \ \ \ \ \ \ \ \ \ \ \ \ + \mu|\varphi|^2 + \lambda |\varphi|^4
\end{eqnarray} 
where $\varphi$ is the field of the composite bosons, $a_{\mu}$ is the Chern-Simons gauge field, $\A$ is a
vector potential for the  external magnetic field and $A_0= V_{imp}$ is the impurity potential.
The grand state of the fractional quantum Hall systems is 
characterized as Bose condensed
state of the composites.
In the dual description, the vortices condensed state describes the 
Hall insulator  phase.
 Blok and Wen extended it to the hierarchical quantum Hall 
states.\cite{bw1,bw2}
 In the case of $\nu=N/(2mN\pm1)$, 
the dual form Lagrangian is given as 
   
\bea
  {\cal L}[b_{I\mu},j_{I\ vortex}^{\mu}]
  &=&\frac{1}{4\pi}K_{IJ} \epsilon^{\mu\nu\rho}b_{I\mu}\partial_{\nu}b_{J\rho}
    \nonumber \\
   &&-\frac{1}{g_I}f_{I\mu\nu}f_I^{\mu\nu}  
   +b_{I\mu}j_{I\ vortex}^{\mu}   \nonumber \\  .
\eea
where $K$ is the K-matrix\cite{wz}, $ j_{vortex}^{\mu}$
is the current of the vortices,\cite{leefisher} and dual gauge field $b_{\mu}$ is 
defined as $J^{\mu}=\frac{1}{2\pi}\varepsilon^{\mu\nu\rho}\partial_{\nu}b_{\rho}$,
$J_{\mu}$ is the electric current vector of the composite bosons. 
For simplicity we consider the case $\nu=N/(2mN+1)\equiv \nu_0(N,m)$ in this paper.
 The extension to 
$\nu=N/(2mN-1)$ is straightfoward.

 The vortices  are quasi-particles in the quantum Hall liquids.
 Increasing the magnetic field, the vortices are introduced and are trapped by the impurity 
potential.  Assuming the impurity potential varies slowly,
 many vortices gather around the maximum points of the potential. 
We can regard their regions as 
void of the quantum Hall liquids. Namely there are two kind of regions 
in the systems, one is filled and the other is empty with the
 quantum Hall liquids. 
The quantum Hall liquids have the gapless edge modes on the boundaries
 of their systems.\cite{wen1,wen2,wen3}
If the impurity potential exists, the edge modes exist not only at 
the sample edges but also on the boundaries between
full and empty regions in the bulk of the sample.
In the case the number of the vortices is small, since the area
 of the empty space is also small,
 these inner edge modes hardly contribute to the 
transport and the transport coefficients are determined by
 the edge modes of the sample boundary
which is connected with the source and the drain electrodes. 
Increasing the number of vortices, however, 
inter-boundaries  edge tunneling occur frequently and they affect
 the transports.

Since the vortex excitations are gapfull, we neglect their dynamics.
Instead we consider the gapless edge modes and their tunneling.  
To see the dynamics of the edge modes, we 
call the contour at the $i$-th boundary between the filled and empty regions
as $C_i$, and take curved cordinate $s_i$ along $C_i$ and coordinate $y_i$
perpendicular to $C_i$.
One can assume that the charge and current densities near 
the $C_i$  decay exponentially
from $C_i$ to quantum Hall liquids side as\cite{nk}  
\bea
 \rho(s,y,t) &=& e^{\lambda y}\rho_{1D}(s,t)  \\ 
 J_s(s,y,t) &=& e^{\lambda y} J_{1D}(s,t) .
\eea

Substituting them to (2) and integrating $y$ we have the Euclidean action;
\be
  S_0 = \sum_i\oint_{C_i} ds d\tau \  \frac{{\rm i}}{4\pi}K_{IJ} \frac{\partial
  \phi_I^i}{\partial \tau} \frac{\partial \phi_J^i}{\partial s}
            + \frac{1}{4\pi}v_{IJ} \frac{\partial \phi_I^i}{\partial s}
            \frac{ \partial \phi_J^i}{\partial s}  \nonumber \\
 \ee
where
the phase fields $\phi^i_I$ is defined as $\rho_{1D}=(1/2 \pi )\partial\phi/\partial s$,
$J_{1D}=-(1/2\pi)\partial \phi/\partial t$ and
 we assume that the distances of $C_i$ and  $C_j$ are longer than $1/\lambda$.
On the points where these distances are comparable to $1/\lambda$, we have to consider the quasi-particle tunneling.
It is described by 
\be
 S_1 = -\sum_{ij,\alpha} \int d\tau \ \  u_{IJ}({\bf x_{\alpha}})\  \cos(\phi_I^i({\bf x_{\alpha}}) -\phi_J^j
  ({\bf x}_{\alpha})),
\ee
where ${\bf x}_{\alpha}$'s are the positions where tunneling occur.
This model is generalized Chalker-Coddington model.
 As well as the $\nu>1$ integer quantum Hall state, there are
some quasi-particles tunneling processes in the hierarchical
state.
  We consider only most relevant processes in the meaning of the
scaling theory which are dominant in the low energy physics.
Imura and Nagaosa studied the scaling theory for the single 
point-contact systems with two edge channels.\cite{imna}
 From their results,
we can find that  the most relevant processes are the tunneling 
between the same edge channels. ($i.e. \ I=J$ in eq.(6) )
We study the case $u_{IJ}=u_I \delta_{IJ}$.
  
 First we consider the weak coupling limit. In this case all
 circuits are widely
separated and quasi-particle tunneling scarcely occurs and do not affect the transport.
 Therefore the weak coupling limit of $S=S_0+S_1$ describes the quantum Hall liquid phase.
Because vortices are trapped by the impurity potential, the transport properties
 are determined by only edge modes on the sample boundary.
The two-terminal conductance is given as $G= \nu(N,m)e^2/h $.
         
Next, to study the properties of the strong coupling regime,
 we derive 
the effective action 
which includes only the non-linear degrees of the freedoms in
eq.(5) and (6).
\cite{kf,fn}
\bea
  Z&=& \int{ \cal D} \phi \ \exp(-S_0[\phi]-S_1[\phi^i_I -\phi^j_I])   \nonumber \\
   &=& \int{ \cal D} \phi \int{ \cal D} \theta \  
 \prod_{ij,\alpha}  \delta(\theta_I({\bf x}_{\alpha})- 
\phi^i_I({\bf x}_{\alpha})+ \phi^j_I({\bf x}_{\alpha}))\nonumber \\
 &&\ \ \ \ \  \ \exp(-S_0[\phi_I^i]-S_1[\theta({\bf x}_{\alpha})])
\nonumber \\ 
  &=& \int{ \cal D} \phi \int{ \cal D} \theta  \int{ \cal D} \lambda \nonumber \\
&&\ \ \ \ \ \ \ \ \ \exp(-S_0[\phi]-S_1[\theta] \nonumber \\
&&  \  \  \ \ \ \ -\sum_{ij,\alpha}  \int d\tau \lambda_I( {\bf x}_{\alpha})
(\theta_I({\bf x}_{\alpha})- 
\phi^i_I({\bf x}_{\alpha})- \phi^j_I({\bf x}_{\alpha})) \nonumber \\
&=& \int{ \cal D} \theta  \exp(-S_{eff}[\theta])
\eea
where
\bea
 S_{eff}[\theta_{\alpha}]
&=& \sum_{\omega}\ \frac{|\omega|}{4\pi}
M_{\alpha \beta} 
{\theta}_{I}({\bf x}_{\alpha},-\omega)K_{IJ}
\theta_{J}({\bf x}_{\beta},\omega)
  \nonumber \\
&&  \ \ \ \ \ \
    -\sum_{I,\alpha} \int d\tau\  u_I(\bf x_{\alpha})\ 
\cos (\theta_{I}({\bf x}_{\alpha},\tau)) \nonumber \\
\eea
$\theta_{I}({\bf x}_{\alpha}) = \phi_I^i({\bf x}_{\alpha})-\phi_I^j({\bf x}_{\alpha})$.
The matrix $M_{\alpha \beta}$ depends on the geometry of the
 distribution of the empty and the filled region. Pryadko and Chaltikian. 
calculated this matrix in very simple cases at 
$\nu = 1/q$.\cite{pryadko}
If  all the tunneling points are widely separated from each other, 
we can regard $M_{\alpha \beta}$ as 
Kronecker's delta $\delta_{\alpha \beta}$.
According to Caldeira and Legget, the diagonal terms of (8) are 
friction terms
 of quantum mechanical
particles.\cite{cl}  In the periodic potential, Bloch states are
 formed and $\theta$ can 
take arbitrary values.
 However the frictions make these particles practically localized,
 but slightly tunnel between the minima of 
the periodic potentials.
In the strong coupling limit, $\theta_I$'s are fixed to 
$2\pi n$ 
where $n$ is an integer.
Because the edge currents are given as the time differential 
of the $\phi$, all quasi-particles tunnel at the tunneling points.

\vspace{5mm}

\vspace{3mm}
{\bf {3. Dual transformation}}
\vspace{3mm}

 To study the dynamics of (8) in the strong 
coupling regime, we introduce the dilute instanton gas approximation.\cite{schmid}
We consider the number of instantons $n$ and their
 configurations as the trace of the partition function;

\bea
  Z &=& Tr \ {\rm e}^{-\beta H} \nonumber \\
    &=& \sum_{\alpha,I} \sum_{n=1}^{\infty} \sum_{\{e_i^{I\alpha}\}} \tilde{u}_I^n({\bf x}_{\beta})
\ \int_0^{\beta} d\tau_n^{I\alpha} \int_0^{\tau_n^{I \alpha}} d\tau_{n-1}^{I\alpha} 
\nonumber \\ &&\ \ \ \ \ \ \ \ \ \ \ \ \ \ \ \ \ \ \ \ \ \ \ \ \ \ \ \ \cdots 
\int_0^{\tau_2^{I\alpha}} d\tau_1^{I\alpha} \ {\rm e}^{-S_{D}}
\eea  
where $\tilde{u}_I({\bf x}_{\beta})\sim e^{-S_{ins}}$ 
is the fugacity of the instantons, 
  $S_{ins}$ is the action of  the single instanton.

\bea
  S_{D} &=&  \sum_{\omega} \frac{|\omega|}{4\pi}K_{IJ}M_{\alpha \beta}
\frac{1}{\omega^2}\sum_{ij}e_i^{I\alpha}e_j^{J\beta} \nonumber \\
&& \ \ \ \ \ \ \ \ \ \ \ \ \ \ \ \   \times  h_{I\alpha}(-\omega)\nonumber    h_{J\beta}(\omega)
{\rm e}^{ {\rm i} \omega(\tau_i^{I\alpha} - \tau_j^{J\beta})}
 \nonumber \\
 &\cong& \sum_{ij} ( \frac{\pi}{\beta} K_{IJ}\sum_{\omega}\frac{1}{|\omega|} 
{\rm e}^{ {\rm i} \omega(\tau_i^{I\alpha} - \tau_j^{J\beta})})e_i^{I\alpha}M_{\alpha\beta}\ e_j^{J\beta} 
\nonumber \\
\eea
 $h_{I\alpha}(\tau)$ is the differential of 
the instanton solution which has the step around $\tau=0$,
  $h_{I\alpha}(\omega)$
is its Fourier transform and
$e_i^{I\alpha}$ is the charge of the instantons.

Using Stratonovich-Hubbard transformation we have

\bea
{\rm e}^{-S_{D}} &=& \int {\cal D}q_{I\alpha}(\tau) 
\nonumber \\ && \ \ \ \ \  \exp (   -\frac{1}{4\pi} K_{IJ}^{-1} \sum_{\omega}
|\omega| q_{I\alpha}(-\omega)M_{\alpha\beta}^{-1}q_{J\beta}(\omega)  \nonumber \\
 && \ \ \ \ \ \ \ \ \ \ \ \ \ \ \ \ \ \  + \sum_{\omega} \frac{\rm i}{ \sqrt{\beta}}\sum_i e_i^{I\alpha}
{\rm e}^{ -{\rm i} \omega \tau_i^{I\alpha}         } q_{I\alpha}(\omega) ).
\nonumber \\
\eea

Substituting (11) for (9), we get the dual representation of (8).

\bea
Z = \int {\cal D} q_{I\alpha} \  {\rm e}^{-\tilde{S}}
\eea

\bea
{\tilde{S}} &=& \sum_{\omega} \frac{|\omega|}{4\pi} M_{\alpha \beta}^{-1} q_{I\alpha}(-\omega)K_{IJ}^{-1} q_{J\beta}(\omega)
\nonumber \\
 &&\ \ \ \ \ + \sum_{I\alpha}\int d\tau \tilde {u}_{I}({\bf x}_{\alpha})
 \cos( q_{I\alpha}) 
\nonumber \\  
  &=&  \sum_{\omega} \frac{|\omega|}{4\pi} M_{\alpha \beta}^{-1} {\tilde {\theta}}_{I}({\bf x}_{\alpha},-\omega)
K_{IJ} {\tilde 
{\theta}}_{J}({\bf x}_{\beta},\omega)
\nonumber \\
 &&\ \ \ \ \ + \sum_{I\alpha}\int d\tau \ {\tilde {u}}_{I}({\bf {x}}_{\alpha}) \cos( K_{IJ}
{\tilde {\theta}}_{J}({\bf x}_{\alpha},\tau)) 
\eea
where ${\tilde {\theta}}_I=K_{IJ}^{-1}q_J$ is the dual field of
  $\theta_I$. 
The first term of (13) is equivalent to the first term of (8)
 except the difference of $M_{\alpha\beta}$
and $M_{\alpha\beta}^{-1}$. Contrarily, the second term of
 (8) describes the tunneling of the quasi-particles,
the second term of (13) describes the electron tunneling.

Eq.(13) describes the quantum droplets junction systems with  weak electron tunneling. This accords with 
 our intuitive picture of the Hall insulator.\cite{simshoni} Actually the  changing $M$ into $M^{-1}$ shows the 
changing of the geometry between the quasi-particle tunneling picture of QH liquid phase and the 
electron tunneling picture of Hall insulator phase.

\vspace{5mm}
{\bf {4. Plateau transition}}
\vspace{3mm}

Now let us consider the effects of  randomness. We read the 
randomness strength as the number of the minima 
and maxima of the impurity potential.
 In the case the value of the magnetic field is just 
$B = B_0(N,m)$ and the randomness strength is  weak enough,
 everywhere of the sample is filled by the quantum 
Hall liquids and there  are no vortices and no edge modes 
in the bulk of the sample.
 If the impurity potential is dominant rather than the 
electron-electron interaction, the quantum Hall
liquids are broken.
Next we consider the case the magnetic field is slightly increased.
 The vortices are introduced to the sample
and pinned by the potential. Then many empty regions exist 
and many edge modes are formed. Assuming
the randomness is   weak enough we can expect to apply the weak coupling edge state network model. We 
consider what happen as the randomness is increased. 
As increase of the number of the minima and maxima,
the number of the empty regions and the tunneling points are 
also increased. So the quasi-particles 
tunneling becomes more frequent.
Therefore in the framework of the edge state network model, 
increasing the randomness strength is regarded effectively
as increasing of the tunneling amplitudes $u({\bf x}_{\alpha})$.
Increasing the magnetic field instead of the randomness strength,
 the area of the empty space is increased and the tunneling 
amplitudes are also increased.
 As we stated above, 
the values of the tunneling amplitudes $u_I$ depend on the channel $I$ which 
corresponds to the Landau levels
of the composite fermions in the Jain's theory.\cite{jain}
 On every tunneling points, the amplitudes of the highest 
Landau level $u_N$ are  largest,  $u_{N-1}$ is the next, 
and so on.
Therefore increasing $\{u_I\}$, the $I=N$ edge modes are pinned first, and the next is $I=N-1$.
To see  the transport coefficients, let us consider a two 
terminal Hall bar. 
For simplify we consider the two terminal conductance. 
The discussion of the two terminal conductance is replaced
as problem of the percolation as following.

 In the weak tunneling limit, the transport properties are
 determined by the charged edge modes theory of sample boundary 
$\phi_{c} = \sum_{I=1}^N \phi_I/N$
which is connected with source and drain electrodes and has the
Luttinger liquid parameter given as 
$K_c = \nu_0(N,m)/N=  1/(2Nm+1) $.
Using linear response theory, we have the two-terminal 
conductance $G = K_c N e^2/h= \frac{N}{2Nm+1}e^2/h$.
Increasing the tunneling amplitudes, the modes $I=N$ 
is pinned  and edge quasi-particles on the boundary of
 the sample are
carried into the  bulk and reach the other side of the 
boundary of the sample. Finally, they return to the 
source electrode, $i.e.$ the $I=N$ channel do not contribute
 to the  transport.
In this case,  the charged  mode is redefined as 
$\phi_{c\pm} = \sum_{I=1}^{N-1}
 \phi_I/(N-1)$ 
and the Luttinger parameter is $K_c = \nu_0(N-1,m)/(N-1)$.
So the two terminal conductance changes from  
$G = \frac{N}{2Nm+1}e^2/h$ to $G = K_c (N-1)e^2/h= 
\frac{N-1}{2(N-1)m+1}e^2/h$.
Thus  the plateaus transition is described as {\it {the 
percolation problem}}.

In the above arguments, we assume that the impurity potential 
$V_{imp}$ varies slowly compared with the magnetic length $l_B$.
Increasing randomness, the ratio $|\nab V_{imp}/V_{imp}| $ 
is increased.
In the case of  $|\nab V_{imp}/V_{imp}| \sim l_B^{-1}$, we can expect 
that the tunneling amplitudes $u_I$ scarcely 
depend on the channel index $I$. We can understand this in composite 
fermion picture as following.
The each channels of composite fermions near the Fermi level are 
very close. Then if the value of the magnetic field
is changed, the quantum Hall liquid phase at $\nu= N/(2Nm+1)$ may 
translate to the Hall insulator phase.

\vspace{5mm}

{\bf {5. Global phase diagram}}
\vspace{3mm}

Now we discuss  the global phase diagram.
We first consider the regime that the randomness is weak. 
If the value of the magnetic field is just 
$B= B_0(N,m)\equiv 2\pi \rho_e/ \nu_0(N,m)$,
the system is filled by quantum Hall liquid and doesn't matter how 
  randomness changes.
If the value of the magnetic field is shifted as $\Delta B =B -B_0(N,m) > 0$,
we can introduce the edge state network model as we said above. If $\Delta B$ is small, the quasi-particle tunneling 
is rare.
Since increasing $\Delta B$ is regarded as increase of the tunneling
 amplitudes $u_I$ in 
the network model, the quantum Hall liquids are translated to the other quantum Hall liquid according 
the selection rules as $\Delta B$ increase.  This statement is consistent with  
changing of the filling factor in Jain's
hierarchical theory. This is the  behavior of 
the global phase diagram when 
  randomness is fixed and the magnetic field is changed. 
Next we consider changing the randomness with fixed magnetic field.
 In the case of 
$\Delta B=0$, the quantum Hall state is stable against weak randomness as a perturbation.
  Thus increasing the randomness, nothing goes on.
If $V_{imp}$ is more dominant than the $e^2/4\pi l_B$ 
which is the characteristic energy of general quantum 
Hall liquids, the condensation of the composite boson is 
broken and the system becomes the Anderson insulator.
Then the plateaus transition does not occur as increasing 
randomness in $B= B_0(N,m)$.
Because the percolation is not allowed, this result is also
 applicable to the case of 
$B  \stackrel{<}{_\sim} B_0(N,m)$. 
In the other case $i.e.$ $B $ is larger 
than $ B_0(N,m)$ but smaller than $B_0(N-1,m)$,
 as randomness increases  from the weak randomness regime, 
quasi-particle tunneling becomes frequent and the quantum Hall 
liquid-quantum Hall liquid transition occurs. 
Because we set the value of the magnetic field smaller than 
$B_0(N-1,m)$ and the filling factor becomes $\nu_0(N-1,m)$ 
effectively,
 the additional transition to the quantum Hall liquid
at $\nu_0(N-2,m)$ doesn't occur. 

These results are consistent with 
an interesting experiment by Reznikov et.al.\cite{rez}
They observed a fractional charge $e/5$ in the shot noise experiment at $\nu=2/5$ constricted quantum Hall liquid.
The two-terminal conductance shows a transition from a plateau at $G=(2/5)e^2/h$ to another plateau at $G=(1/3)e^2/h$
 as the constriction is increase. On the second plateau they observed a current carrying particle with charge $e/3$.
Motivated by the experiment K.Imura and one of the authors(K.N) calculated the shot noise at $G=(2/5)e^2/h$ regime and 
$G=(1/3)e^2/h$ regime in the framework of the chiral Tomonaga-Luttinger liquid theory  in the single point contact
systems.\cite{in}
The edge state network model is regarded as a generalization of  above analysis. 

\vspace{5mm}
{\bf {6. Conclusion}}
\vspace{3mm}

In summary, we introduced a general filling version of the
  edge state network model which has the 
dual description 
and discussed the transition between different quantum Hall liquid 
sates. The results are following.

 i) If $B$ is almost $B_0(N,m)$ or smaller than $B_0(N,m)$, 
  the plateau transition from $\nu= \nu_0(N,m)$ quantum  
Hall liquid phase doesn't occur as   randomness increases.

 ii) If $B$ is enough larger than $B_0(N,m)$, 
$\nu=\nu_0(N,m)$ quantum Hall liquid shows 
a plateau transition to 
$\nu = \nu_0(N-1,m)$ quantum Hall liquid phase.
 However  additional transition doesn't occur.
The condition for the plateau transition have been 
discussed in 
the section 4.

So far we have discussed the case $\nu=N/(2mN+1)$. 
To construct the 
effective theory for the $\nu=N/(2mN-1)$ 
quantum Hall state,
one should change the sign of the charge of the partons in 
the theory of the parton construction.
Eventually we should regard decrease of the magnetic field as
increase  of that in above theory.
 These statements  lead to a broad form of the global
 phase diagram  shown as Fig.1
which seems like Kravechenko's one. \cite{ex2}
If randomness is weak, the plateau to plateau transitions
 occur as  the magnetic field is increased, or decreased
 for $\nu=N/(2mN+1)$, and  ($\nu=N/(2mN-1)$), respectively.
Contrary, in the regime where randomness is strong to some degree,
the transition from a plateau is not to the other plateau
 but to the insulator. Consequently,
the striking point which differs from KLZ's
diagram is that the hierarchical regimes 
border upon the Hall insulating regime.  
We expect that if one study the transport properties near the transition point of 
the hierarchical quantum Hall liquid states to the Hall insulating states, charge-flux duality symmetry is 
observed as the experiments by Shahar et. al.\cite{shahar1,shahar2,simshoni}

\end{multicols}
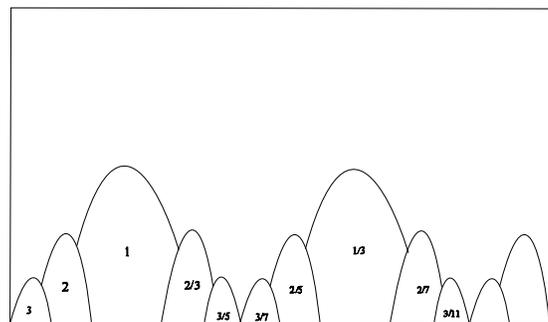
\end{document}